\documentclass[12pt,letterpaper]{article}
\usepackage{amsmath,amsfonts,amsthm,amssymb}

\theoremstyle{plain}

\numberwithin{equation}{section}
\newtheorem{thm}{Theorem}[section]
\newtheorem{lem}[thm]{Lemma}
\newtheorem{cor}[thm]{Corollary}

\newcounter{cond}

\newcommand{\complex}{{\mathbb C}}
\newcommand{\real}{{\mathbb R}}
\newcommand{\ascript}{{\mathcal A}}
\newcommand{\bscript}{{\mathcal B}}
\newcommand{\dscript}{{\mathcal D}}
\newcommand{\escript}{{\mathcal E}}
\newcommand{\fscript}{{\mathcal F}}
\newcommand{\qscript}{{\mathcal Q}}
\newcommand{\rmre}{\mathrm{Re\,}}
\newcommand{\rmspan}{\mathrm{span}}
\newcommand{\rmrank}{\mathrm{rank}}
\newcommand{\rmrange}{\mathrm{Range}}

\newcommand{\ab}[1]{\left|#1\right|}
\newcommand{\doubleab}[1]{\left\|#1\right\|}
\newcommand{\brac}[1]{\left\{#1\right\}}
\newcommand{\paren}[1]{\left(#1\right)}
\newcommand{\sqbrac}[1]{\left[#1\right]}
\newcommand{\elbows}[1]{{\left\langle#1\right\rangle}}
\newcommand{\ket}[1]{{\left|#1\right>}}
\newcommand{\bra}[1]{{\left<#1\right|}}

\begin{document}

\title{HILBERT SPACE REPRESENTATIONS\\
OF DECOHERENCE FUNCTIONALS\\
AND QUANTUM MEASURES}
\author{Stan Gudder\\ Department of Mathematics\\
University of Denver\\ Denver, Colorado 80208\\
sgudder@math.du.edu}
\date{}
\maketitle

\begin{abstract}
We show that any decoherence functional $D$ can be represented by a spanning vector-valued measure on a complex Hilbert space. Moreover, this representation is unique up to an isomorphism when the system is finite. We consider the natural map $U$ from the history Hilbert space $K$ to the standard Hilbert space $H$ of the usual quantum formulation. We show that $U$ is an isomorphism from $K$ onto a closed subspace of $H$ and that $U$ is an isomorphism from $K$ onto $H$ if and only if the representation is spanning. We then apply this work to show that a quantum measure has a Hilbert space representation if and only if it is strongly positive. We also discuss classical decoherence functionals, operator-valued measures and quantum operator measures.
\end{abstract}

\section{Introduction}  % Section 1
In the usual quantum description of a physical system, we begin with a complex Hilbert space $H$. The states of the system are represented by density operators, the observables by self-adjoint operators and the dynamics by unitary operators on $H$. In the history approach to quantum mechanics and in applications such as quantum gravity and cosmology, one defines a useful concept called a decoherence functional $D$ \cite{djso10, gudamm, hal09, sor94}. It is believed by researchers in these fields that $D$ encodes important information about the system. For example, $D$ can be employed to find the interference between quantum objects and can also be used to find a quantum measure that quantifies the propensity that quantum events occur \cite{dgt08, gudms, gud09, sor94}.

Because of the fundamental importance of $D$, it appears to be useful to reverse this formalism. We propose to begin with a decoherence functional $D$ with natural properties and to then reconstruct the usual quantum formulation. We consider two types of reconstruction that we call vector and operator representations of $D$. We show that there always exists a spanning vector representation of $D$ and when the system is finite, this representation is unique up to an isomorphism. For a finite system, cyclic operator representations always exist but for infinite systems, their existence is unknown.

Besides the standard Hilbert space $H$ of the usual quantum formulation, there exists a history Hilbert space $K$ that is directly associated with $D$ \cite{djso10}. Moreover, we can define a natural map $U\colon K\to H$
\cite{djso10}. We show that $U$ is an isomorphism from $K$ onto a closed subspace of $H$ and that $U$ is an isomorphism from $K$ onto $H$ if and only if the vector representation is spanning.

We also present several characterizations of classical decoherence functionals. We show that a quantum measure has a Hilbert space representation if and only if it is strongly positive. We briefly consider quantum operator measures generated by decoherence operators.

\section{Vector Representations} % Section 2
Let $(\Omega ,\ascript )$ be a measurable space. The elements of $\Omega$ represent outcomes and the sets in the $\sigma$-algebra $\ascript$ represent events for a physical system or process. A
\textit{decoherence functional} $D\colon\ascript\times\ascript\to\complex$ from the Cartesian product of $\ascript$ with itself into the complex numbers satisfies the following conditions \cite{djso10, hal09, sor07}:

\begin{list} {(D\arabic{cond})}{\usecounter{cond}
\setlength{\rightmargin}{\leftmargin}}
%(D1)
\item $D(\Omega ,\Omega )=1$,
%(D2)
\item $A\mapsto D(A,B)$ is a complex measure for all $B\in\ascript$.
%(D3)
\item If $A_1,\ldots ,A_n\in\ascript$, then $D(A_i,A_j)$ is a positive semi-definite $n\times n$ matrix.
\end{list}
Condition~(D1) is an inessential normalization property that does not affect any of the results in this paper. Notice that (D3) implies $D(A,A)\ge 0$ and $D(A,B)=\overline{D(B,A)}$.

We now give two examples of decoherence functionals. If $\nu\colon\ascript\to\complex$ is a complex measure satisfying $\nu (\Omega )=1$, we can view $\nu$ as an amplitude measure for a physical system. It is easy to check that $D\colon\ascript\times\ascript\to\complex$ given by $D(A,B)=\nu (A)\overline{\nu (B)}$ is a decoherence functional. The map $\mu\colon\ascript\to\real ^+$ given by
\begin{equation}         % equation (2.1)
\label{eq21}
\mu (A)=D(A,A)=\ab{\nu (A)}^2
\end{equation}
is an example of a quantum measure \cite{dgt08, gudms, gud09, sor94, sor07} and these will be treated in Section~6. This is an example of a vector representation of $D$.

The second example is more general and illustrates an operator representation of $D$. Let $H$ be a complex Hilbert and denote the set of bounded linear operators from $H$ to $H$ by $B(H)$. We say that
$\escript\colon\ascript\to B(H)$ is an \textit{operator-valued measure} if for every sequence of mutually disjoint sets $A_i\in\ascript$ and every $\phi ,\phi '\in H$ we have
\begin{equation*}
\elbows{\escript (\cup A_i)\phi ,\phi '}=\sum\elbows{\escript (A_i)\phi ,\phi '}
\end{equation*}
where the summation converges absolutely. If $\escript\colon\ascript\to B(H)$ is an operator-valued measure and
$\psi\in H$ is a unit vector, we define $D\colon\ascript\times\ascript\to\complex$ by

\begin{equation}         % equation (2.2)
\label{eq22}
D(A,B)=\elbows{\escript (A)\psi ,\escript (B)\psi}
\end{equation}
If $D(\Omega ,\Omega )=1$, then it is easy to check that $D$ is a decoherence functional. If the closed span
\begin{equation*}
\overline{\rmspan}\brac{\escript (A)\psi\colon A\in\ascript}=H
\end{equation*}
we say that $\psi$ is a \textit{cyclic vector} for $\escript$. Again, the map $\mu\colon\ascript\to\real ^+$ defined by
\begin{equation}         % equation (2.3)
\label{eq23}
\mu (A)=D(A,A)=\doubleab{\escript (A)\psi}^2
\end{equation}
is an example of a quantum measure.

\begin{lem}       % Lemma 2.1
\label{lem21}
If $D$ is an $n\times n$ positive semi-definite matrix, then there exists a complex Hilbert space $H$ and a spanning set of vectors $e_i\in H$, $i=1,\ldots ,n$, such that $D_{ij}=\elbows{e_i,e_j}$. Also, if $D_{ij}=\elbows{f_i,f_j}$ for a spanning set of vectors $f_i$ in a complex Hilbert space $K$, then there is a unitary operator $U\colon H\to K$ such that $Ue_i=f_i$, $i=1,\ldots ,n$.
\end{lem}
\begin{proof}
Since $D$ is positive semi-definite, the map
\begin{equation*}
\elbows{f,g}=\sum D_{ij}f(i)\overline{g(j)}
\end{equation*}
becomes an indefinite inner product on the vector space $\complex ^n$. Defining $\doubleab{f}=\elbows{f,f}^{1/2}$, let $N\subseteq\complex ^n$ be the subspace
\begin{equation*}
N=\brac{f\in\complex ^n\colon\doubleab{f}=0}
\end{equation*}
Letting $H$ be the quotient space $H=\complex ^n/N$, the elements of $H$ become $[f]=f+N$, $f\in\complex ^n$. Then 
$H$ is a finite-dimensional complex Hilbert space with inner product $\elbows{[f],[g]}=\elbows{f,g}$. Letting
$e_1,\ldots ,e_n$ be the standard basis for $\complex ^n$ we have that
\begin{equation*}
\elbows{[e_i],[e_j]}=\sum D_{rs}e_i(r)\overline{e_j(s)}=D_{ij}
\end{equation*}
Since $\brac{e_1,\ldots ,e_n}$ spans $\complex ^n$, $\brac{[e_1],\ldots ,[e_n]}$ spans $H$. We can assume without loss of generality that $\brac{[e_1],\ldots ,[e_m]}$ forms a basis for $H$, $m\le n$. Then
\begin{equation*}
\dim H=m=n-\dim N=\rmrank (D)
\end{equation*}
Now suppose that $D_{ij}=\elbows{f_i,f_j}$ for a spanning set of vectors $f_i\in K$, $i=1,\ldots ,n$. It is well-known that 
$\rmrank (D)$ is the number of linearly independent rows of $D$. Since $\brac{[e_1],\ldots ,[e_m]}$ are linearly independent we have that the first $m$ rows of $D$ are linearly independent. We now show that $f_1,\ldots ,f_m$ are linearly independent. Suppose that $\sum _{i=1}^m\alpha _if_i=0$ for $\alpha _i\in\complex$. Then
$\sum _{i=1}^m\alpha _i\elbows{f_i,f_j}=0$ for $j=1,\ldots ,n$, and hence,
\begin{equation*}
\alpha _1\paren{\elbows{f_1,f_1},\ldots ,\elbows{f_1,f_n}}
  +\cdots +\alpha _m\paren{f_m,f_1},\ldots ,\elbows{f_m,f_n}=0
\end{equation*}
We conclude that $\alpha _1,\ldots ,\alpha _m=0$ so $f_1,\ldots ,f_m$ are linearly independent. It follows that
$f_1,\ldots ,f_m$ form a basis for $K$. Define the operator $U\colon H\to K$ by $U[e_i]=f_i$, $i=1,\ldots ,m$, and extend by linearity. We then have that
\begin{equation*}
\elbows{U[e_i],U[e_j]}=\elbows{f_i,f_j}=D_{ij}=\elbows{[e_i],[e_j]}
\end{equation*}
$i=1,\ldots ,m$. Since any $[f]\in H$ has a unique representation
\begin{equation*}
[f]=\sum _{i=1}^m\alpha _i[e_i]
\end{equation*}
we have that
\begin{align*}
\doubleab{U[f]}^2&=\elbows{\sum\alpha _iU[e_i],\sum\alpha _jU[e_j]}
  =\sum\alpha _i\overline{\alpha _j}\elbows{U[e_i],U[e_j]}\\
  &=\sum\alpha _i\overline{\alpha _j}\elbows{[e_i],[e_j]}=\elbows{[f],[f]}=\doubleab{f}^2
\end{align*}
Since $U$ is surjective, $U$ is unitary.
\end{proof}

A map $\escript\colon\ascript\to H$ is a \textit{vector-valued measure} on $H$ if for any sequence of mutually disjoint
sets $A_i\in\ascript$ we have that
\begin{equation*}
\lim _{n\to\infty}\sum _{i=1}^n\escript (A_i)=\escript (\cup A_i)
\end{equation*}
in the norm topology. A \textit{vector representation} for a decoherence functional
$D\colon\ascript\times\ascript\to\complex$ is a pair $(H,\escript )$ where $\escript\colon\ascript\to H$ is a
vector-valued measure satisfying
\begin{equation}         % equation (2.4)
\label{eq24}
D(A,B)=\elbows{\escript (A),\escript (B)}
\end{equation}
for all $A,B\in\ascript$. If $\overline{\rmspan}\brac{\escript (A)\colon A\in\ascript}=H$, then $(H,\escript )$ is called a
\textit{spanning vector representation} for $D$. If $\Omega =\brac{\omega _1,\ldots ,\omega _n}$, then we let
$\ascript =2^\Omega$ and call $(\Omega ,\ascript )$ a \textit{finite measurable space}. It is clear that any map
$D\colon\ascript\times\ascript\to\complex$ satisfying $D(\Omega ,\Omega )=1$ and \eqref{eq24} is a decoherence functional. The next two results show that the converse holds.

\begin{thm}       % Theorem 2.2
\label{thm22}
If $(\Omega ,\ascript )$ is a finite measurable space and $D\colon\ascript\times\ascript\to\complex$ is a decoherence functional, then there exists a spanning vector representation $(H,\escript )$ for $D$. Moreover, if $(K,\fscript )$ is a spanning vector representation for $D$, then there is a unitary operator $U\colon H\to K$ such that
$U\escript (A)=\fscript (A)$ for every $A\in\ascript$.
\end{thm}
\begin{proof}
Since $D$ is a decoherence functional, we have that $D_{ij}=D(\omega _i,\omega _j)$ is positive semi-definite. By Lemma~\ref{lem21}, there exists a spanning set $e_1,\ldots ,e_n$ in a Hilbert space $H$ such that
$D_{ij}=\elbows{e_i,e_j}$. For $A\in\ascript$, define $\escript\colon\ascript\to H$ by
\begin{equation*}
\escript (A)=\sum\brac{e_i\colon\omega _i\in A}
\end{equation*}
Then $\escript$ is a vector-valued measure and we have
\begin{align*}
D(A,B)&=\sum\brac{D(\omega _i,\omega _j)\colon\omega _i\in A,\omega _j\in B}
  =\sum _{ij}\brac{\elbows{e_i,e_j}\colon\omega _i\in A,\omega _j\in B}\\
  &=\elbows{\sum\brac{e_i\colon\omega _i\in A},\sum\brac{e_j\colon\omega _j\in B}}
  =\elbows{\escript (A),\escript (B)}
\end{align*}
Hence $(H,\escript )$ is a spanning vector representation of $D$. For the second statement of the theorem, let
$e_i=\escript (\omega _i)$, $f_i=\fscript (\omega _i)$, $i=1,\ldots ,n$. It is clear that $\rmspan\brac{e_1,\ldots ,e _n}=H$ and similarly $\rmspan\brac{f_1,\ldots ,f_n}=K$. By Lemma~\ref{lem21}, there is a unitary operator $U\colon H\to K$ such that $Ue_i=f_i$. Therefore,
\begin{align*}
U\escript (A)&=U\sqbrac{\sum\brac{e_i\colon\omega _i\in A}}=\sum\brac{Ue_i\colon\omega _i\in A}\\
  &=\sum\brac{f_i\colon\omega _i\in A}=\fscript (A)
\end{align*}
for all $A\in\ascript$.
\end{proof}

For an arbitrary measurable space $(\Omega ,\ascript )$, we cannot use the method in the proof of
Theorem~\ref{thm22}. Moreover, we do not know whether the uniqueness result in Theorem~\ref{thm22} holds in general.

\begin{thm}       % Theorem 2.3
\label{thm23}
If $(\Omega ,\ascript )$ is a measurable space and $D\colon\ascript\times\ascript\to\complex$ is a decoherence functional, then there exists a spanning vector representation $(H,\escript )$ for $D$.
\end{thm}
\begin{proof}
Let $S$ be the set of all complex-valued measurable functions on $\Omega$ with a finite number of values (simple functions). Any $f\in S$ has a canonical representation $f=\sum a_i\chi _{A_i}$ where $a_i\ne a_j$, $A_i\cap A_j=\emptyset$ for $i\ne j$ and $a_i\ne 0$, $i,j=1,\ldots ,n$. If $f=\sum a_i\chi _{A_i}$, $g=\sum b_j\chi _{B_j}$ are canonical representations, we define
\begin{equation}         % equation (2.5)
\label{eq25}
\elbows{f,g}=\sum _{i,j}a_i\overline{b_j}D(A_i,B_j)
\end{equation}
It is straightforward to show that \eqref{eq25} holds even if the representations of $f$ and $g$ are not canonical. It is also easy to verify that $\elbows{\cdot\,,\cdot}$ is an indefinite inner product. As in Lemma~\ref{lem21}, we let $N$ be the subspace of $S$ given by
\begin{equation*}
N=\brac{f\in S\colon\|f\|=0}
\end{equation*}
Letting $H_0=S/N$, the elements of $H_0$ are the equivalence classes $[f]=f+N$, $f\in S$. We define the inner product $\elbows{\cdot\,,\cdot}$ on $H_0$ by $\elbows{[f],[g]}=\elbows{f,g}$. Letting $H$ be the completion of $H_0$ we have that $H_0$ is a dense subspace of the HIlbert space $H$. Defining $\escript\colon\ascript\to H$ by
$\escript (A)=[\chi _A]$ we have that
\begin{align*}
&\overline{\rmspan}\brac{\escript (A)\colon A\in\ascript }=H\\
\intertext{and}
&\elbows{\escript (A),\escript (B)}=\elbows{\chi _A,\chi _B}=D(A,B)
\end{align*}
To show that $\escript$ is a vector-valued measure, let $A_i\in\ascript$ be mutually disjoint, $i=1,2,\ldots\,$. We then have that
\begin{align*}
\left\|\escript\paren{\cup A_i}\right.&\left.-\sum _{i=1}^n\escript (A_i)\right\|^2\\
  &=\doubleab{\escript\paren{\cup A_i}}^2+\doubleab{\sum _{i=1}^n\escript (A_i)}^2
  -2\rmre\elbows{\escript\paren{\cup A_i},\sum _{i=1}^n\escript (A_i)}\\
  &=D(\cup A_i,\cup A_i)+\sum _{i,j=1}^nD(A_i,A_j)-2\rmre\sum _{i=1}^nD(\cup A_i,A_i)
\end{align*}
Applying Condition~(D2) we conclude that
\begin{equation*}
\lim _{n\to\infty}\sum _{i=1}^n\escript (A_i)=\escript (\cup A_i)
\end{equation*}
in the norm topology.
\end{proof}

Results similar to Theorems~\ref{thm22} and \ref{thm23} have appeared in \cite{djsu10}.

\section{Operator Representations} % Section 3
An \textit{operator representation} for a decoherence functional $D\colon\ascript\times\ascript\to\complex$ is a triple
$(H,\escript ,\psi )$ where $H$ is a complex Hilbert space, $\psi\in H$ is a unit vector and
$\escript\colon\ascript\to B(H)$ is an operator-valued measure such that \eqref{eq22} holds for every $A,B\in\ascript$. We say that $(H,\escript ,\psi )$ is \textit{cyclic} if $\psi$ is a cyclic vector for $\escript$. We call $\escript (A)$ the
\textit{event} or \textit{class operator} at $A$. It is not hard to show that if $(H,\escript ,\psi )$ is an operator representation for $D$, then $\fscript (A)=\escript (A)\psi$ gives a vector representation for $D$. However, the operator representation gives more information because it specifies the class operator at every $A\in\ascript$. Moreover, we do not know whether every vector representation $(H,\fscript )$ has a corresponding operator representation
$(H,\escript ,\psi )$ such that $\fscript (A)=\escript (A)\psi$ for all $A\in\ascript$. Two operator representations
$(H,\escript ,\psi )$ and $(K,\fscript ,\phi )$ are \textit{equivalent} if there exists a unitary operator $U\colon H\to K$ such that $U\psi =\phi$ and $U\escript (A)U^*=\fscript (A)$ for all $A\in\ascript$. For example, if $(H,\escript ,\psi )$ is an operator representation for $D$ and $\alpha\in\complex$ with $\ab{\alpha}=1$, then $(H,\escript ,\alpha\psi )$ is an equivalent operator representation for $D$. In this case, the unitary operator is $U=\alpha I$.

We shall show that a decoherence functional on a finite measurable space possesses an operator representation. It is an open problem whether this result holds for an arbitrary decoherence functional. It should be pointed out that although finiteness is a strong restriction, there are important applications for finite quantum systems. For example, models for quantum computation and information are usually finite. Moreover, measurement based quantum computation has a structure that is similar to that of the history approach to quantum mechanics \cite{joz09}.

\begin{thm}       % Theorem 3.1
\label{thm31}
If $(\Omega ,\ascript )$ is a finite measurable space and $D\colon\ascript\times\ascript\to\complex$ is a decoherence functional, then there exists a cyclic operator representation for $D$.
\end{thm}
\begin{proof}
By Lemma~\ref{lem21}, there exists a spanning set $e_i,\ldots ,e_n$ in a Hilbert space $H$ such that
$D(\omega _i,\omega _j)=\elbows{e_i,e_j}$, $i,j=1,\ldots ,n$. We show by induction on $m$ that there is a $\phi\in H$ such that $\elbows{e_i,\phi}\ne 0$ for all $e_i\ne 0$, $i=1,\ldots ,m\le n$. The result clearly holds for $m=1$. Assume that the result holds for $m$. Then there is a $\phi$ such that $\elbows{e_i,\phi}\ne 0$ for all $e_i\ne 0$,
$i=1,\ldots ,m$. Suppose $e_{m+1}\ne 0$ and $\elbows{e_{m+1},\phi}=0$. By continuity, we can find a small ball $B\subseteq H$ centered at $\phi$ such that $\elbows{e_i,f}\ne 0$ for all $f\in B$ and $e_i\ne 0$, $i=1,\ldots ,m$. If $\elbows{e_{m+1},f}=0$ for all $f\in B$ then $e_{m+1}=0$ which is a contradiction. Hence, there is an $f\in B$ such that $\elbows{e_i,f}\ne 0$ for all $e_i\ne 0$, $i=1,\ldots ,m+1$. This completes the induction proof. Letting
$\psi=\phi/\|\phi\|$ we conclude that $\psi\in H$ is a unit vector satisfying $\elbows{e_i,\psi}\ne 0$ for all $e_i\ne 0$, $i=1,\ldots ,n$. Define $P_i\in B(H)$, $i=1,\ldots ,n$, as follows. If $e_i=0$, then $P_i=0$ and if $e_i\ne 0$, then
\begin{equation*}
P_i=\frac{1}{\elbows{e_i,\psi}}\,\ket{e_i}\bra{e_i}
\end{equation*}
We then have that
\begin{equation*}
\elbows{P_i\psi ,P_j\psi}=D(\omega _i,\omega _j)
\end{equation*}
for $i,j=1,\ldots n$. Defining $\escript\colon\ascript\to B(H)$ by
\begin{equation*}
\escript (\ascript )=\sum\brac{P_i\colon\omega _i\in\ascript}
\end{equation*}
we have that $(H,\escript ,\psi )$ is a cyclic operator representation for $D$.
\end{proof}

We now give an example which shows that there may exist inequivalent cyclic operator representations for $D$. Let
$\Omega =\brac{1,2}$ and let $D\colon 2^\Omega\times 2^\Omega\to\complex$ be the decoherence functional given by $D(\emptyset ,A)=D(A,\emptyset )=0$, $D(\Omega ,\Omega )=1$
\begin{align*}
D(i,j)&=\frac{1}{5}\,
\left[\begin{matrix}\noalign{\smallskip}1&1\\\noalign{\smallskip}1&2\\\noalign{\smallskip}\end{matrix}\right]\\
  D(\Omega ,1)&=D(1,\Omega )=2/5\\\noalign{\smallskip}
  D(\Omega ,2)&=D(2,\Omega )=3/5
\end{align*}
Let $H=\complex ^2$ with the usual inner product and standard basis $e_1,e_2$. Define the operator-valued measure $\fscript\colon 2^\Omega\to H$ by $\fscript (\emptyset )=0$, $\fscript (1)=c\ket{e_1}\bra{e_1}$,
$\fscript (2)=cI$ and
\begin{equation*}
\fscript (\Omega )=c\ket{e_1}\bra{e_1}+cI
\end{equation*}
where $c=\sqrt{2/5\,}$. Let $\phi$ be the unit vector $\phi =2^{-1/2}(1,1)$. Since
$\fscript (1)\phi =\frac{c}{\sqrt{2\,}}\,e_1$ and $\fscript (2)\phi =c\phi$ we see that $\phi$ is cyclic for $\fscript$. Moreover,
\begin{align*}
\elbows{\fscript (1)\phi ,\fscript (1)\phi}&=\frac{c^2}{2}=\frac{1}{5}=D(1,1)\\\noalign{\smallskip}
\elbows{\fscript (2),\phi ,\fscript (2)\phi}&=c^2=\frac{2}{5}=D(2,2)\\\noalign{\smallskip}
\elbows{\fscript (1)\phi ,\fscript (2)\phi}&=\frac{c^2}{\sqrt{2\,}}\,\elbows{e_1,\phi}
  =\frac{c^2}{2}=\frac{1}{5}=D(1,2)=D(2,1)
\end{align*}
It follows that $\elbows{\fscript (A),\fscript (B)}=D(A,B)$ for all $A,B\in 2^\Omega$. We conclude that
$(H,\fscript ,\phi )$ is a cyclic operator representation for $D$. Since $\rmrank\paren{\fscript (2)}=2$ and
$\rmrank\paren{\escript (2)}=1$ where $\escript (2)$ is the operator defined in Theorem~\ref{thm31},
$(H,\fscript  ,\phi )$ is not equivalent to $(H,\escript ,\psi )$ of Theorem~\ref{thm31}.

\section{History Hilbert Space} % Section 4
Let $D\colon\ascript\times\ascript\to\complex$ be a decoherence functional and $K_0$ the set of complex-valued functions on $\ascript$ that vanish except for a finite number of sets in $\ascript$. For $f,g\in K_0$ define
\begin{equation*}
\elbows{f,g}=\sum _{A,B\in\ascript}D(A,B)f(A)\overline{g(B)}
\end{equation*}
As before, we define the subspace
\begin{equation*}
N=\brac{f\in K_0\colon\|f\|=0}
\end{equation*}
The quotient space $K_1=K_0/N$ consists of equivalence classes $[f]=f+N$, $f\in K_0$. Again
$\elbows{[f],[g]}=\elbows{f,g}$ becomes an inner product on $K_1$. We denote the completion of $K_1$ by $K$ and call $K$ the \textit{history Hilbert space} for $D$ \cite{djso10}. The space $K$ corresponds to the history approach to quantum mechanics
\cite{hal09, ish94, sor07}.

Let $(H,\escript )$ be a vector representation for $D$. We think of $H$ as the standard Hilbert space of the usual quantum formulation.  A natural connection between $K$ and $H$ was introduced in \cite{dgt08}. We define the
\textit{natural map} $U\colon K_0\to H$ by
\begin{equation*}
Uf=\sum _{A\in\ascript}f(A)\escript (A)
\end{equation*}
It is clear that $U$ is linear and moreover,
\begin{align*}
\elbows{Uf,Ug}&=\elbows{\sum _{A\in\ascript}F(A)\escript (A),\sum _{B\in\ascript}g(B)\escript (B)}\\
&=\sum _{A,B\in\ascript}f(A)\overline{g(B)}\elbows{\escript (A),\escript (B)}\\
&=\sum _{A,B\in\ascript}f(A)\overline{g(B)}D(A,B)=\elbows{f,g}
\end{align*}
Hence, $U\colon K_1\to H$ given by $U[f]=Uf$ is well-defined and is an isometry. It follows that $U$ has a unique extension to an isometry, that we also denote by $U$, from $K$ into $H$. The next result shows that $K$ is isomorphic to a closed subspace of $H$ and characterizes when $K$ is isomorphic to all of $H$. This proves a conjecture posed in \cite{djso10}.

\begin{thm}       % Theorem 4.1
\label{thm41}
The operator $P=UU^*$ is an orthogonal projection on $H$ and $U\colon K\to PH$ is unitary. The natural map
$U\colon K\to H$ is unitary if and only if $(H,\escript )$ is spanning.
\end{thm}
\begin{proof}
The operator $P$ is clearly self-adjoint and since $U^*U=I_K$ we have that
\begin{equation*}
P^2=UU^*UU^*=UU^*=P
\end{equation*}
Clearly $PH\subseteq\rmrange (U)$. Conversely, if $\phi\in\rmrange (U)$ then $\phi =U\phi '$ for some $\phi '\in K$. Again, $U^*U=I_K$ gives
\begin{equation*}
P\phi =PU\phi '=UU^*U\phi =U\phi '=\phi
\end{equation*}
Hence, $PH=\rmrange (U)$. Thus, $U\colon K\to PH$ is unitary from $K$ to the closed subspace $PH$ of $H$. Now it is clear that
\begin{equation*}
\overline{\rmspan}\brac{\escript (A)\colon (A\in\ascript}=\rmrange (U)
\end{equation*}
Hence, $\rmrange (U)=H$ if and only if $\escript$ is spanning. It follows that $U\colon K\to H$ is unitary if and only if
$(H,\escript )$ is spanning.
\end{proof}

We can proceed in a similar way for an operator representation $(H,\escript ,\psi )$ for $D$. Then the corresponding vector representation $(H,\fscript )$ given by $\fscript (A)=\escript (A)\psi$ is spanning if and only if $(H,\escript ,\psi )$ is cyclic. By Theorem~\ref{thm41} the natural map $U\colon K\to H$ given by
\begin{equation}         % equation (4.1)
\label{eq41}
Uf=\sum _{A\in\ascript}f(A)\escript (A)\psi
\end{equation}
is unitary if and only if $(H,\escript )$ is cyclic.

We now introduce an example presented in \cite{djso10}. Consider a system consisting of a single particle that has
$n$ possible positions $\brac{1,2,\ldots ,n}$ at any time. We assume that the particle evolves in $N-1$ discrete time steps at times $0=t_1<t_2<\cdots <t_N=T$. Each history $\omega$ of the system is represented by an $N$-tuple of integers $\omega =(\omega _1,\ldots ,\omega _N)$ with $1\le\omega _i\le n$, $i=1,\ldots ,N$, where $\omega _i$ is the location of the particle at time $t_i$. The corresponding sample space $\Omega$ is the collection of
$n^N$ possible histories and $\ascript =2^\Omega=\brac{A\colon A\subseteq\Omega}$. For this example, the standard Hilbert space is $H=\complex ^n$ with the usual inner product
\begin{equation*}
\elbows{\phi ,\phi '}=\sum _{i=1}^n\phi _i\overline{\phi '_i}
\end{equation*}
where $\phi =(\phi _1,\ldots ,\phi _n)$. The initial state is given by a fixed unit vector $\psi\in H$.

To describe the decoherence functional, we assume that states propagate from time $t$ to time $t'$ according to a unitary evolution operator $U(t',t)$ that satisfies
\begin{equation*}
U(t'',t')U(t',t)=U(t'',t)
\end{equation*}
Let $P_1,\ldots ,P_n$ be the projection operators given by
\begin{equation*}
P_i(\phi _1,\ldots ,\phi _n)=(0\ldots ,0,\phi _i,0,\ldots ,0)
\end{equation*}
$i=1,\ldots ,n$. These projections form the spectral measure for the position operator. For a path
$\omega =(\omega _1,\ldots ,\omega _N)$ we define the path operator
\begin{equation}         % equation (4.2)
\label{eq42}
\escript (\omega )=P_{\omega _N}U(t_N,t_{N-1})P_{\omega _{N-1}}
  \cdots P_{\omega _3}U(t_3,t_2)P_{\omega _2}U(t_2,t_1)P_{\omega _1}
\end{equation}
We next define the event operator (or class operator) $\escript (A)$, $A\in\ascript$, by
\begin{equation*}
\escript (A)=\sum _{\omega\in A}\escript (\omega )
\end{equation*}
Then $\escript\colon\ascript\to L(H)$ becomes an operator-valued measure and $(H,\escript ,\psi )$ is an operator representation for the decoherence functional $D\colon\ascript\times\ascript\to\complex$ given by
\begin{equation*}
D(A,B)=\elbows{\escript (A)\psi ,\escript (B)\psi}
\end{equation*}

So far we have presented the standard quantum formulation for the system. We now construct the history Hilbert space $K$ for the decoherence functional $D$ just defined. We have seen that the natural map $U\colon K\to H$ given by \eqref{eq41} is an isometry from $K$ into $H$. Theorem~\ref{thm41} tells us that $U$ is unitary if and only if
$(H,\escript ,\psi )$ is cyclic. Another sufficient condition for $U$ to be unitary is given in \cite{djso10}. We now show that this condition is also necessary.

\begin{thm}       % Theorem 4.2
\label{thm42}
For this example, $U$ is unitary if and only if for every $i=1,\ldots ,n$ there exists an $\omega\in\Omega$ such that
\begin{equation}         % equation (4.3)
\label{eq43}
\sqbrac{\escript (\omega )\psi}(i)\ne 0
\end{equation}
\end{thm}
\begin{proof}
Let $\brac{\psi ^1,\ldots ,\psi ^n}$ be the standard basis for $\complex ^n$. By \eqref{eq42} we have that
$\escript (\omega )\psi =c(\omega )\psi ^{\omega _N}$ for some $c(\omega )\in\complex$. If \eqref{eq43} holds, then
$\escript (\omega )\psi =c(\omega )\psi ^i$ for $c(\omega )\ne 0$. It follows that $\psi$ is cyclic so by
Theorem~\ref{thm41}, $U$ is unitary. Conversely, suppose $\sqbrac{\escript (\omega )\psi}(i_0)=0$ for every
$\omega\in\Omega$. It follows that if
\begin{equation*}
\phi\in\rmspan\brac{\escript (A)\psi\colon A\in\ascript}
\end{equation*}
then $\phi (i_0)=0$. Hence, $\psi$ is not cyclic so by Theorem~\ref{thm41}, $U$ is not unitary.
\end{proof}

\section{Classical Decoherence Functionals} % Section 5
A decoherence functional $D\colon\ascript\times\ascript\to\complex$ is \textit{weakly classical} if
$\mu (A)=D(A,A)$ is a probability measure on $\ascript$. A decoherence functional
$D\colon\ascript\times\ascript\to\complex$ is \textit{classical} if $D(A,B)=\mu (A\cap B)$ for some probability measure on $\ascript$. Of course, $D$ is weakly classical if $D$ is classical.

\begin{thm}       % Theorem 5.1
\label{thm51}
{\rm (a)}\enspace A decoherence functional $D\colon\ascript\times\ascript\to\complex$ is weakly classical if and only if there exists a probability measure $\mu\colon\ascript\to\real ^+$ such that $\rmre D(A,B)=\mu (A\cap B)$.
{\rm (b)}\enspace If $D\colon\ascript\times\ascript\to\complex$ has the form $D(A,B)=\mu (A\cap B)$ for some probability measure $\mu\colon\ascript\to\real ^+$, then $D$ is a classical decoherence functional.
\end{thm}
\begin{proof}
(a)\enspace If $\rmre D(A,B)=\mu (A\cap B)$ for some probability measure $\mu$, it is clear that $D$ is weakly classical. Conversely, suppose $D$ is weakly classical so that $\mu (A)=D(A,A)$ is a probability measure. By Theorem~\ref{thm23}, there is spanning vector representation $(H,\escript )$ so that $D(A,B)=\elbows{\escript (A),\escript (B)}$. If $A,B\in\ascript$ are disjoint, then
\begin{align*}
\elbows{\escript (A),\escript (A)}&+\elbows{\escript (B),\escript (B)}=\mu (A)+\mu (B)=\mu (A\cup B)\\
  &=\elbows{\escript(A\cup B),\escript (A\cup B}=\elbows{\escript (A)+\escript (B),\escript (A)+\escript (B)}\\
  &=\elbows{\escript (A),\escript (A)}+\elbows{\escript (B),\escript (B)}+2\rmre\elbows{\escript (A),\escript (B)}
\end{align*}
Hence,
\begin{equation*}
\rmre D(A,B)=\rmre\elbows{\escript (A),\escript (B)}=0
\end{equation*}
For arbitrary $A,B\in\ascript$ we have
\begin{align*}
\rmre D(A,B)&=\rmre D\sqbrac{(A\cap B)\cup (A\cap B'),(A\cap B)\cup (A'\cap B)}\\
  &=\rmre\left[D(A\cap B,A\cap B)+D(A\cap B,A'\cap B)\right.\\
  &\qquad\left.+D(A\cap B',A\cap B)+D(A\cap B',A'\cap B)\right]\\
  &=\rmre D(A\cap B,A\cap B)=\mu (A\cap B)
\end{align*}
(b)\enspace Suppose $D(A,B)=\mu (A\cap B)$ for a probability measure $\mu\colon\ascript\to\real ^+$. We only need to show that $D$ is a decoherence functional. It is clear that $D(\Omega ,\Omega )=1$ and that $A\mapsto D(A,B)$ is a complex measure for every $B\in\ascript$. Let $A_1,\ldots ,A_k\in\ascript$ and let $\ascript _0$ be the Boolean algebra generated by $\brac{A_1,\ldots ,A_k}$. Since $\ab{\ascript _0}<\infty$, by Stone's theorem there is a finite set
$\Omega =\brac{\omega _1,\ldots ,\omega _n}$ and an isomorphism $h\colon 2^\Omega\to\ascript _0$. Define
$D'\colon 2^\Omega\times 2^\Omega\to\complex$ by $D'(A,B)=D\paren{h(A),h(B)}$. In particular,
\begin{equation*}
D'_{ij}=D'(\omega _i,\omega _j)=D\paren{h(\omega _i),h(\omega _j)}
\end{equation*}
Now, $\sum\limits _{i,j}D'_{ij}=1$ and for $i\ne j$ we have
\begin{equation*}
D'_{ij}=D\paren{h(\omega _i),h(\omega _j)}=\mu\paren{h(\omega _i)\cap h(\omega _i)}=0
\end{equation*}
Hence, $D'(\omega _i,\omega _j)=\mu\paren{h(\omega _i)}\delta _{ij}$, $i,j=1,\ldots ,n$ so $\sqbrac{D'_{ij}}$ is a positive semi-definite matrix. It follows from the proof of Theorem~\ref{thm22} that there exists a vector representation $(H,\escript )$ such that
\begin{equation*}
D'(\omega _i,\omega _j)=\elbows{\escript (\omega _i),\escript (\omega _j)}
\end{equation*}
for $i,j=1,\ldots ,n$. Hence, $D'(A,B)=\elbows{\escript (A),\escript (B)}$ so $D'$ is a decoherence functional. Hence,
\begin{equation*}
D(A_i,A_j)=D'\paren{h^{-1}(A_i),h^{-1}(A_j)}
\end{equation*}
is a positive semi-definite matrix. We conclude that $D$ is a decoherence functional.
\end{proof}

\begin{thm}       % Theorem 5.2
\label{thm52}
If $D\colon\ascript\times\ascript\to\complex$ is a decoherence functional, the following statements are equivalent.
{\rm (a)}\enspace $D$ is classical.
{\rm (b)}\enspace $D(A\cap B,C)=D(B,A\cap C)$ for all $A,B,C\in\ascript$.
{\rm (c)}\enspace If $A\cap B=\emptyset$, then $D(A,B)=0$.
{\rm (d)}\enspace $D$ has a spanning vector representation $(H,\escript )$ where $\escript (A)\perp\escript (B)$ whenever $A\cap B=\emptyset$.
\end{thm}
\begin{proof}
For (a)$\Rightarrow$(b), if $D$ is classical, then $D(A,B)=\mu (A\cap B)$ for a probability measure $\mu\colon\ascript\to\real ^+$. Hence
\begin{equation*}
D(A\cap B,C)=\mu\paren{(A\cap B)\cap C}=\mu\paren{B\cap (A\cap C)}=D(B,A\cap C)
\end{equation*}
For (b)$\Rightarrow$(c), suppose (b) holds and $A\cap B=\emptyset$. We have that
\begin{equation*}
D(A,B)=D(A\cap B,B)=D(\emptyset ,B)=0
\end{equation*}
For (c)$\Rightarrow$(d), suppose (c) holds. By Theorem~\ref{thm23}, $D$ has a spanning vector representation
$(H,\escript )$ such that $D(A,B)=\elbows{\escript (A),\escript (B)}$ for all $A,B\in\ascript$. If $A\cap B=\emptyset$, then $D(A,B)=0$ so that $\escript (A)\perp\escript (B)$.

\noindent For (d)$\Rightarrow$(a), suppose (d) holds. We conclude that
\begin{align*}
D(A,B)&=\elbows{\escript (A),\escript (B)}
  =\elbows{\escript (A\cap B)+\escript (A\cap B'),\escript (A\cap B)+\escript (B\cap A')}\\
  &=\elbows{\escript (A\cap B),\escript (A\cap B)}=\doubleab{\escript (A\cap B)}^2
\end{align*}
Defining $\mu\colon\ascript\to\real ^+$ by $\mu (A)=\doubleab{\escript (A)}^2$ we have that $D(A,B)=\mu (A\cap B)$. To show that $\mu$ is a probability measure, we have
\begin{equation*}
\mu (\Omega )=\mu (\Omega\cap\Omega )=D(\Omega ,\Omega )=1
\end{equation*}
Moreover, if $A_i\in\ascript$ are mutually disjoint, then
\begin{align*}
\mu\paren{\cup A_i}&=\doubleab{\escript\paren{\cup A_i}}^2
  =\doubleab{\lim _{n\to\infty}\sum _{i=1}^n\escript (A_i)}^2
  =\lim _{n\to\infty}\doubleab{\sum _{i=1}^n\escript (A_i)}^2\\
  &=\lim _{n\to\infty}\sum _{i=1}^n\doubleab{\escript (A_i)}^2=\sum _{i=1}^\infty\mu (A_i)\qedhere
\end{align*}
\end{proof}
The importance of Theorem~\ref{thm52} is that it characterizes classical decoherence functionals in terms of their vector representations. In fact, we have the following corollary.

\begin{cor}       % Corollary 5.3
\label{cor53}
A decoherence functional $D\colon\ascript\times\ascript\to\complex$ is classical if and only if for any vector representation $(H,\escript )$ for $D$ we have $\escript (A)\perp\escript (B)$ whenever $A\cap B=\emptyset$.
\end{cor}

The following is another characterization of classicality.

\begin{cor}       % Corollary 5.4
\label{cor54}
A decoherence functional $D$ is classical if and only if\newline
$D(A,B)=D(A\cap B,A\cap B)$ for all $A,B\in\ascript$.
\end{cor}
\begin{proof}
If $D(A,B)=D(A\cap B,A\cap B)$, then $A\cap B=\emptyset$ implies that 
\begin{equation*}
D(A,B)=D(\emptyset, \emptyset )=0
\end{equation*}
By Theorem~\ref{thm52}, $D$ is classical. Conversely, if $D$ is classical by Theorem~\ref{thm52} we have
\begin{equation*}
D(A\cap B,A\cap B)=D(A,A\cap B)=D(A,B)\qedhere
\end{equation*}
\end{proof}

\section{Quantum Measures} % Section 6
This section applies our previous work on decoherence functionals to the study of quantum measures. For
$(\Omega ,\ascript )$ a measurable space, a map $\mu\colon\ascript\to\real ^+$ is \textit{grade}-2
\textit{additive} if
\begin{equation}         % equation (6.1)
\label{eq61}
\mu (A\cup B\cup C)=\mu (A\cup B)+\mu (A\cup C)+\mu (B\cup C)-\mu (A)-\mu (B)-\mu (C)
\end{equation}
for all mutually disjoint $A,B,C\in\ascript$. A $q$-\textit{measure} is a grade-2 additive set function $\mu\colon\ascript\to\real ^*$ that satisfies the following conditions.

\begin{list} {(C\arabic{cond})}{\usecounter{cond}
\setlength{\rightmargin}{\leftmargin}}
%(C1)
\item If $A_1\subseteq A_2\subseteq\cdots$ is an increasing sequence in $\ascript$, then
\begin{equation*}
\lim _{n\to\infty}\mu (A_n)=\mu\paren{\bigcup _{i=1}^\infty A_i}
\end{equation*}
%(C2)
\item If $A_1\supseteq A_2\supseteq\cdots$ is a decreasing sequence in $\ascript$, then
\begin{equation*}
\lim _{n\to\infty}\mu (A_n)=\mu\paren{\bigcap _{i=1}^\infty A_i}
\end{equation*}
\end{list}
Using the notation $A\triangle B=(A\cap B')\cup (A'\cap B)$, it is shown in \cite{dut04} that $\mu\colon\ascript\to\real ^+$ is grade-2 additive if and only if
\begin{equation}         % equation (6.2)
\label{eq62}
\mu (A\cup B)=\mu (A)+\mu (B)-\mu (A\cap B)+\mu (A\triangle B)-\mu (A\cap B')-\mu (A'\cap B)
\end{equation}
for all $A,B\in\ascript$.

Due to quantum interference, a $q$-measure need not satisfy the usual additivity condition of an ordinary measure but satisfies the more general grade-2 additivity condition \eqref{eq61} instead
\cite{gudms, gudamm, hal09, sor94, sor07}. We have already mentioned that \eqref{eq21} and \eqref{eq23} are examples of $q$-measures. If $\mu$ is a $q$-measure on $\ascript$, we call $(\Omega ,\ascript ,\mu )$ a
$q$-\textit{measure space}. We shall not assume that a $q$-measure $\mu$ satisfies $\mu (\Omega )=1$. For this reason we relax Condition~(D1) for a decoherence functional and our previous results still hold.

Let $(\Omega ,\ascript ,\mu )$ be a $q$-measure space in which $\Omega =\brac{\omega _1,\ldots ,\omega _n}$ is finite and $\ascript$ is the power set $2^\Omega$. The \textit{two-point interference term} for $\mu$ is defined by
\begin{equation*}
I_{ij}^\mu=\mu\paren{\brac{\omega _i,\omega _j}}-\mu (\omega _i)-\mu (\omega _j)
\end{equation*}
for $i\ne j=1,\ldots ,n$, where $\mu (\omega _i)=\mu\paren{\brac{\omega _i}}$. The
\textit{decoherence matrix} $D$ is given by
\begin{align*}
D_{ii}&=D(\omega _i,\omega _j)=\mu (\omega _i),\qquad i=1,\ldots ,n\\
D_{ij}&=D(\omega _i,\omega _j)=\tfrac{1}{2}\,I_{ij}^\mu,\qquad i\ne j=1,\ldots ,n
\end{align*}
The $q$-measure $\mu$ is \textit{strongly positive} if $D$ is positive semi-definite. Of course, if $\mu$ is a measure, then $I_{ij}^\mu =0$ for $i\ne j$ so $\mu$ is strongly positive. However, there are many examples of $q$-measures that are not strongly positive. For instance, let $\Omega =\brac{\omega _1,\omega _2}$ and define the $q$-measure
$\mu\colon\ascript\to\real ^+$ by $\mu (\Omega )=1$ and
\begin{equation*}
\mu (\emptyset )=\mu (\omega _1)=\mu (\omega _2)=0
\end{equation*}
Then $\mu$ is not strongly positive because
\begin{equation*}
D=
\left[\begin{matrix}\noalign{\smallskip}0&1\\\noalign{\smallskip}1&0\\\noalign{\smallskip}\end{matrix}\right]
\end{equation*}
is not positive semi-definite. For another example, let $\Omega =\brac{\omega _1,\omega _2,\omega _3}$ and define the $q$-measure $\mu\colon\ascript\to\real ^+$ by $\mu (\emptyset )=\mu (\Omega )=0$ and $\mu (A)=1$ for
$A\ne\emptyset ,\Omega$. Then $\mu$ is not strongly positive because
\begin{equation*}
D=
\left[\begin{matrix}\noalign{\smallskip}\ \ 1&-1&-1\\\noalign{\smallskip}-1&\ \ 1&-1\\
-1&-1&\ \ 1\\\noalign{\smallskip}\end{matrix}\right]
\end{equation*}
is not positive semi-definite. 

\begin{thm}       % Theorem 6.1
\label{thm61}
Let $(\Omega ,\ascript )$ be a finite measurable space. A map $\mu\colon\ascript\to\real ^+$ is a strongly positive
$q$-measure if and only if there exists a finite-dimensional complex Hilbert space $H$ and a spanning vector-valued measure $\escript\colon\ascript\to H$ such that
\begin{equation}         % equation (6.3)
\label{eq63}
\mu (A)=\doubleab{\escript (A)}^2
\end{equation}
for all $A\in\ascript$.
\end{thm}
\begin{proof}
Let $\Omega =\brac{\omega _1,\ldots ,\omega _n}$. It is straightforward to check that if $\mu$ has the form
\eqref{eq63}, then $\mu$ is a strongly positive $q$-measure. Conversely, suppose that
$\mu\colon\ascript\to\real ^+$ is a strongly positive $q$-measure and let $D_{ij}$ be the corresponding positive semi-definite decoherence matrix. By Lemma~\ref{lem21} and the proof of Theorem~\ref{thm22}, there exists a decoherence functional $D\colon\ascript\times\ascript\to\complex$ given by
\begin{equation*}
D(A,B)=\sum\brac{D_{ij}\colon\omega _1\in A,\omega _j\in B}
\end{equation*}
a finite-dimensional complex Hilbert space $H$ and a spanning vector-valued measure $\escript\colon\ascript\to H$ such that
\begin{equation*}
D(A,B)=\elbows{\escript (A),\escript (B)}
\end{equation*}
for all $A,B\in\ascript$. Notice that \eqref{eq63} holds if $A=\brac{\omega _i}$, $i=1,\ldots ,n$. To show that
\eqref{eq63} holds for a general $A\in\ascript$, we can assume without loss of generality that
$A=\brac{\omega _1,\ldots ,\omega _m}$, $2\le m\le n$. It follows from Theorem~2.2 of reference \cite{djso10} that
\begin{equation*}
\mu (A)=\sum _{i<j=1}^m\mu\paren{\brac{\omega _i,\omega _j}}-(m-2)\sum _{i=1}^m\mu (\omega _i)
\end{equation*}
We then have that
\begin{align*}
\doubleab{\escript (A)}^2&=D(A,A)=\sum _{i,j=1}^mD_{ij}=\sum _{i=1}^mD_{ii}+2\sum _{i<j=1}^mD_{ij}\\
&=\sum _{i=1}^m\mu (\omega _i)+\sum _{i<j=1}^mI_{ij}^\mu\\
&=\sum _{i=1}^m\mu (\omega _i)+\sum _{i<j=1}^m\mu\paren{\brac{\omega _i,\omega _j}}
  -\sum _{i<j=1}^m\sqbrac{\mu (\omega _i)-\mu (\omega _j)}\\
&=\sum _{i=1}^m\mu (\omega _i)+\sum _{i<j=1}^m\mu\paren{\brac{\omega _i,\omega _j}}
  -(m-1)\sum _{i=1}^m\mu (\omega _i)\\
&=\sum _{i<j=1}^m\mu\paren{\brac{\omega _i,\omega _j}}-(m-2)\sum _{i=1}^m\mu (\omega _i)
  =\mu (A)\qedhere
\end{align*}
\end{proof}

If the sample space $\Omega$ is infinite, then we must proceed differently than we did for the finite case. For example, when $\Omega$ is infinite the singleton and doubleton subsets may not be measurable (i.e., may not be in 
$\ascript$) and even if they are measurable, they frequently all have measure zero.

Let $(\Omega ,\ascript ,\mu )$ be a $q$-measure space. For $A,B\in\ascript$ define
\begin{equation}         % equation (6.4)
\label{eq64}
\triangle (A,B)=\tfrac{1}{2}\sqbrac{\mu (A\cup B)+\mu (A\cap B)-\mu (A\cap B')-\mu (A'\cap B)}
\end{equation}
Notice that if $\brac{\omega _i}$ and $\brac{\omega _j}$ are measurable, then
\begin{equation*}
\triangle\paren{\brac{\omega _i},\brac{\omega _j}}=D_{ij}
\end{equation*}
so $\triangle (A,B)$ is a generalization of the decoherence matrix. We say that $\mu$ is \textit{strongly positive} if for any $A_1,\ldots ,A_k\in\ascript$, the matrix $\triangle (A_i,A_j)$, $i,j=1,\ldots ,k$ is positive semi-definite. It follows that this definition reduces to the definition of strongly positive in the finite case. Also, observe that if $\mu$ is a measure, then \eqref{eq64} gives $\triangle (A,B)=\mu (A\cap B)$ so $\triangle$ is a classical decoherence functional. Applying Theorem~\ref{thm52}, there exists a vector-valued measure $\escript\colon\ascript\to H$ satisfying
$\escript (A)\perp\escript (B)$ whenever $A\cap B=\emptyset$ such that $\mu (A)=\doubleab{\escript (A)}^2$ for all
$A\in\ascript$. Although the next result generalizes Theorem~\ref{thm61}, we gave an independent proof of
Theorem~\ref{thm61} because the decoherence matrix $D_{ij}$ is physically more intuitive than $\triangle$.

\begin{thm}       % Theorem 6.2
\label{thm62}
Let $(\Omega ,\ascript )$ be a measurable space. A map $\mu\colon\ascript\to\real ^+$ is a strongly positive
$q$-measure if and only if there exists a complex Hilbert space $H$ and a spanning vector-valued measure
$\escript\colon\ascript\to H$ such that \eqref{eq63} holds.
\end{thm}
\begin{proof}
Suppose $\mu$ has the form \eqref{eq63}. It is straightforward to check that $\mu$ is a $q$-measure. To show that
$\mu$ is strongly positive, let $A_1,\ldots ,A_k\in\ascript$. Applying \eqref{eq64} we have that
\begin{align*}
\triangle (A,B)
&=\tfrac{1}{2}\left[\elbows{\escript (A\cup B),\escript (A\cup B)}+\elbows{\escript (A\cap B),\escript (A\cap B)}\right.\\
&\quad \left.-\elbows{\escript (A\cap B'),\escript (A\cap B')}-\elbows{\escript (A'\cap B),\escript (A'\cap B)}\right]\\
&=\tfrac{1}{2}\left[\doubleab{\escript (A\cap B)+\escript (A\cap B')+\escript (A'\cap B)}^2
  +\doubleab{\escript (A\cap B)}^2\right.\\
&\quad \left.-\doubleab{\escript (A\cap B')}^2-\doubleab{\escript (A'\cap B)}^2\right]\\
&=\rmre\left[\doubleab{\escript (A\cap B)}^2+\elbows{\escript (A\cap B),\escript (A\cap B')}\right.\\
&\quad \left. +\elbows{\escript (A\cap B),\escript (A'\cap B)}+\elbows{\escript (A\cap B'),\escript (A'\cap B)}\right]\\
&=\rmre\sqbrac{\elbows{\escript (A\cap B)+\escript (A\cap B'),\escript (A\cap B)+\escript (A'\cap B)}}\\
&=\rmre\elbows{\escript (A),\escript (B)}
\end{align*}
Hence, for $\alpha _1,\ldots ,\alpha _k\in\complex$ we have
\begin{align*}
\sum _{i,j}\triangle (A_i,A_j)\alpha _i\overline{\alpha _j}
&=\sum _{i,j}\rmre\elbows{\escript (A_i),\escript (A_j)}\alpha _i\overline{\alpha _j}\\
&=\rmre\elbows{\sum\alpha _i\escript (A_i),\sum\alpha _j\escript (A_j)}\ge 0
\end{align*}
We conclude that $\triangle (A_i,A_j)$ is a positive semi-definite matrix so $\mu$ is strongly positive.

Conversely, suppose that $\mu$ is a strongly positive $q$-measure. We show that $A\mapsto\triangle (A,B)$ is a complex-valued measure for every $B\in\ascript$. If $A_1,A_2\in\ascript$ are disjoint we have
\begin{align}         % equation (6.5)
\label{eq65}
\triangle (A_1\cup A_2,B)&=\tfrac{1}{2}\left\{\mu\sqbrac{(A_1\cup B)\cup (A_2\cup B)}
  +\mu\sqbrac{(A_1\cap B)\cup (A_2\cap B)}\right.\notag\\
&\quad \left.-\mu\sqbrac{(A_1\cap B')\cup (A_2\cap B')}-\mu (A'_1\cap A'_2\cap B)\right\}
\end{align}
By \eqref{eq62} we have
\begin{align}         % equation (6.6)
\label{eq66}
\mu&\sqbrac{(A_1\cup B)\cup (A_2\cup B)}\notag\\
&=\mu\sqbrac{(A_1\cup B)\triangle (A_2\cup B)}-\mu\sqbrac{(A_1\cup B)\cap (A_2\cup B)'}\notag\\
&\quad -\mu\sqbrac{(A_1\cup B)'\cap (A_2\cap B)}+\mu (A_1\cup B)+\mu (A_2\cup B)\notag\\
&\quad -\mu\sqbrac{(A_1\cup B)\cap (A_2\cup B)}\notag\\
&=\mu\sqbrac{(A_1\cap B')\cup (A_2\cap B')}-\mu (A_1\cap B')-\mu (A_2\cap B')\notag\\
&\quad +(A_1\cup B)+\mu (A_2\cup B)-\mu (B)
\end{align}
Since $\mu$ is grade-2 additive we have
\begin{align}         % equation (6.7)
\label{eq67}
\mu (B)&=\mu\sqbrac{(B\cap A_1)\cup (B\cap A_2)\cup (B\cap A'_1\cap A'_2)}\notag\\
&=\mu\sqbrac{(B\cap A_1)\cup (B\cap A_2)}+\mu\sqbrac{(B\cap A_1)\cup (B\cap A'_1\cap A'_2)}\notag\\
&\quad +\mu\sqbrac{(B\cap A_2)\cup (B\cap A'_1\cap A'_2)}-\mu (B\cap A_1)-\mu (B\cap A_2)\notag\\
&\quad -\mu (B\cap A'_1\cap A'_2)\notag\\
&=\mu\sqbrac{(B\cap A_1)\cup (B\cap A_2)}+\mu (B\cap A'_2)+\mu (B\cap A'_1)\notag\\
&\quad -\mu (B\cap A_1)-\mu (B\cap A_2)-\mu (B\cap A'_1\cap A'_2)
\end{align}
Substituting \eqref{eq67} into \eqref{eq66} gives
\begin{align}         % equation (6.8)
\label{eq68}
\mu&\sqbrac{(A_1\cup B)\cup (A_2\cup B)}\notag\\
&=\mu\sqbrac{(A_1\cap B')\cup (A_2\cap B')}
-\mu (A_1\cap B')-\mu (A_2\cap B')\notag\\
&\quad +\mu (A_1\cup B)+\mu (A_2\cup B)-\mu\sqbrac{(B\cap A_1)\cup (B\cap A_2)}-\mu (B\cap A'_2)\notag\\
&\quad -\mu (B\cap A'_1)+\mu (B\cap A_1)+\mu (B\cap A_2)+\mu (B\cap A'_1\cap A'_2)
\end{align}
Substituting \eqref{eq68} into \eqref{eq65} gives
\begin{align*}
\triangle (A_1\cup A_2,B)
&=\tfrac{1}{2}\left[\mu (A_1\cup B)+\mu (A_2\cup B)+\mu (A_1\cap B)+\mu (A_2\cap B)\right.\\
&\quad \left.-\mu (A_1\cap B')-\mu (A_2\cap B')-\mu (A'_2\cap B)-\mu (A'_1\cap B)\right]\\
&=\triangle (A_1,B)+\triangle (A_2,B)
\end{align*}
We conclude by induction that
\begin{equation*}
\triangle\paren{\bigcup _{i=1}^nA_i,B}=\sum _{i=1}^n\triangle (A_i,B)
\end{equation*}
whenever $A_1,\ldots ,A_n\in\ascript$ are mutually disjoint. Let $A_i\in\ascript$ with
$A_1\subseteq A_2\subseteq\cdots$. Since $\mu$ is continuous, we have
\begin{align*}
\lim&\triangle (A_i,B)\\
&=\tfrac{1}{2}\sqbrac{\lim\mu (A_i\cup B)+\lim\mu (A_i\cap B)-\lim\mu (A_i\cap B')-\lim (A'_i\cap B)}\\
&=\tfrac{1}{2}\brac{\mu\sqbrac{(\cup A_i)\cup B}+\mu\sqbrac{(\cup A_i)\cap B}-\mu\sqbrac{(\cup A_i)\cap B'}
  -\mu\sqbrac{(\cap A_i)'\cap B}}\\
  &=\triangle (\cup A_i,B)
\end{align*}
It follows that $A\mapsto D(A,B)$ is a complex-valued measure for all $B\in\bscript$. Hence, $D$ is a decoherence functional (except for Condition~(D1)) and the result follows from Theorem~\ref{thm23}
\end{proof}

\section{Operator Quantum Measures} % Section 7
This section briefly considers a generalization of $q$-measures to operator $q$-measures. Let $(\Omega ,\ascript )$ be a measurable space and $\escript\colon\ascript\to B(H)$ be an operator-valued measure. We define the
\textit{decoherence operator} $\dscript\colon\ascript\times\ascript\to B(H)$ by
\begin{equation*}
\dscript (A,B)=\escript (B)^*\escript (A)
\end{equation*}
Notice that if $\doubleab{\escript (\Omega )\psi}=1$, then $D(A,B)=\elbows{\dscript (A,B)\psi ,\psi}$ is a decoherence functional. We call $\qscript\colon\ascript\to B(H)$ given by
\begin{equation}         % equation (7.1)
\label{eq71}
\qscript (A)=\dscript (A,A)=\escript (A)^*\escript (A)
\end{equation}
an \textit{operator} $q$-\textit{measure}. The next result summarizes some of the interesting properties of $\qscript$.

\begin{thm}       % Theorem 7.1
\label{thm71}
If $\escript\colon\ascript\to B(H)$ is an operator-valued measure, then the operator $q$-measure \eqref{eq71} is a positive operator-valued function that satisfies the following conditions.
{\rm (a)}\enspace (\textit{Grade}-2 \textit{additivity}) For any mutually disjoint sets $A,B,C\in\ascript$ we have
\begin{equation*}
\qscript (A\cup B\cup C)=\qscript (A\cup B)+\qscript (A\cup C)+\qscript (B\cup C)-\qscript (A)-\qscript (B)-\qscript (C)
\end{equation*}
{\rm (b)}\enspace (\textit{Regularity}) If $\qscript (A)=0$, then $\qscript (A\cup B)=\qscript (B)$ whenever
$A\cap B=\emptyset$. If $A\cap B=\emptyset$ and $\qscript (A\cup B)=0$, then $\qscript (A)=\qscript (B)$.

\noindent{\rm (c)}\enspace (\textit{Continuity}) If $A_1\subseteq A_2\subseteq\cdots$ and $\phi ,\phi '\in H$, then
\begin{equation*}
\elbows{\qscript (\cup A_i)\phi ,\phi '}=\lim\elbows{\qscript (A_i)\phi ,\phi '}
\end{equation*}
and if $A_1\supseteq A_2\supseteq\cdots$ and $\phi ,\phi '\in H$, then
\begin{equation*}
\elbows{\qscript (\cap A_i)\phi ,\phi '}=\lim\elbows{\qscript (A_i)\phi ,\phi '}
\end{equation*}
\end{thm}
\begin{proof}
It is clear that $\qscript (A)$ is a positive operator for all $A\in\ascript$.
(a)\enspace Since $\qscript (A)=\escript (A)^*\escript (A)$, $A\in\ascript$, we have
\begin{align*}
\qscript (A&\cup B)+\qscript (A\cup C)+\qscript (B\cup C)-\qscript (A)-\qscript (B)-\qscript (C)\\
&=2\escript (A)^*\escript (A)+2\escript (B)^*\escript (B)+2\escript (C)^*\escript (C)+\escript (A)^*\escript (B)\\
  &\quad +\escript (B)^*\escript (A)+\escript (A)^*\escript (C)+\escript (C)^*\escript (A)
  +\escript (B)^*\escript (C)+\escript (C)^*\escript (B)\\
  &\quad -\escript (A)^*\escript (A)-\escript (B)^*\escript (B)-\escript (C)^*\escript (C)\\
  &=\escript (A\cup B\cup C)^*\escript (A\cup B\cup C)=\qscript (A\cup B\cup C)
\end{align*}
(b)\enspace If $\qscript (A)=0$, then $\escript (A)^*\escript (A)=0$. Hence, for every $\phi\in H$ we have
\begin{equation*}
\doubleab{\escript (A)\phi}^2=\elbows{\escript (A)\phi ,\escript (A)\phi}
  =\elbows{\escript (A)^*\escript (A)\phi ,\phi}=0
\end{equation*}
Hence, $\escript (A)\phi =0$ so $\escript (A)=0$. For $A,B\in\ascript$ with $A\cap B=\emptyset$ we have
\begin{align*}
\qscript (A\cup B)&=\escript (A\cup B)^*\escript (A\cup B)=\sqbrac{\escript (A)+\escript (B)}^*
  \sqbrac{\escript (A)+\escript (B)}\\
&=\escript (B)^*\escript (B)=\qscript (B)
\end{align*}
If $A\cap B=\emptyset$ and $\qscript (A\cup B)=0$, then
\begin{align*}
0=\qscript (A\cup B)
&=\escript (A)^*\escript (A)+\escript (B)^*\escript (B)+\escript (A)^*\escript (B)+\escript (B)^*\escript (A)\\
&=\sqbrac{\escript (A)+\escript (B)}^*\sqbrac{\escript (A)+\escript (B)}
\end{align*}
As before, $\escript (A)+\escript (B)=0$. It follows that
\begin{align*}
\escript (B)^*\escript (A)&=-\escript (B)^*\escript (B)\\
\intertext{and}
\escript (A)^*\escript (B)&=-\escript (B)^*\escript (B)
\end{align*}
Hence
\begin{equation*}
\escript (A)^*\escript (A)-\escript (B)^*\escript (B)=0
\end{equation*}
so that $\qscript (A)=\qscript (B)$.

\noindent (c)\enspace Let $A_1\subseteq A_2\subseteq\cdots$ be increasing in $\ascript$ and let $\phi ,\phi '\in H$. Define
$B_1=A_1$, $B_i=A_i\smallsetminus A_{i-1}$ $i=2,3,\ldots\,$. Then $B_i\in\ascript$ are mutually disjoint so we have
\begin{align*}
\elbows{\qscript (\cup A_i)\phi ,\phi '}&=\elbows{\escript (\cup B_i)\phi ,\escript (\cup B_j)\phi '}
  =\sum _{i,j}\elbows{\escript (B_i)\phi ,\escript (B_j)\phi '}\\
  &=\lim _{n,m\to\infty}\elbows{\escript\paren{\bigcup _{i=1}^nB_i}\phi ,\escript\paren{\bigcup _{j=1}^mB_j}\phi '}\\
  &=\lim _{n,m\to\infty}\elbows{\escript (A_n)\phi ,\escript (A_m)\phi '}\\
  &=\lim _{n\to\infty}\elbows{\escript (A_n)^*\escript (A_n)\phi ,\phi '}
  =\lim\elbows{\qscript (A_n)\phi ,\phi '}
\end{align*}
The result is similar for $A_1\supseteq A_2\supseteq\cdots\,$.
\end{proof}

Motivated by Section~5 we make the following definitions. A decoherence operator $\dscript$ is
\textit{classical} if $\dscript (A,B)=0$ whenever $A\cap B=\emptyset$. For an operator $T\in B(H)$ we define
$\rmre T=\frac{1}{2}(T+T^*)$. A decoherence operator $\dscript$ is \textit{weakly classical} if $\rmre\dscript (A,B)=0$ whenever $A\cap B=\emptyset$.

\begin{thm}       % Theorem 7.2
\label{thm72}
Let $\escript\colon\ascript\to B(H)$ be an operator-valued measure and let $\dscript (A,B)=\escript (B)^*\escript (A)$ and $\qscript (A)=\dscript (A,A)$ be the corresponding decoherence operator and operator $q$-measure.
{\rm (a)}\enspace $\dscript$ is classical if and only if $\dscript (A,B)=\qscript (A\cap B)$ for every $A,B\in\ascript$.
{\rm (b)}\enspace $\dscript$ is weakly classical if and only if $\qscript$ is an operator-valued measure.
\end{thm}
\begin{proof}
(a)\enspace If $\dscript (A,B)=\qscript (A\cap B)$ and $A\cap B=\emptyset$, then
\begin{equation*}
\dscript (A,B)=\qscript (\emptyset )=\dscript (\emptyset ,\emptyset )=0
\end{equation*}
so $\dscript$ is classical. Conversely, if $\dscript$ is classical, then
\begin{align*}
\dscript (A,B)&=\escript (B)^*\escript (A)=\sqbrac{\escript (A\cap B)^*+\escript (B\cap A')^*}
\sqbrac{\escript (A\cap B)+\escript (A\cap B')}\\
&=\dscript (A\cap B,A\cap B)+\dscript (A\cap B,A\cap B')+\dscript (B\cap A',A\cap B)\\
&\quad +\dscript (B\cap A',A\cap B')\\
&=\dscript (A\cap B,A\cap B)=\qscript (A\cap B)
\end{align*}
(b)\enspace If $\qscript$ is an operator-valued measure and $A\cap B=\emptyset$, then
\begin{equation*}
\escript (A\cup B)^*\escript (A\cup B)=\qscript (A\cup B)=\qscript (A)+\qscript (B)=\escript (A)^*\escript (A)
+\escript (B)^*\escript (B)
\end{equation*}
Hence,
\begin{equation*}
\rmre\dscript (A,B)=\tfrac{1}{2}\sqbrac{\escript (B)^*\escript (A)+\escript (A)^*\escript (B)}=0
\end{equation*}
so $\dscript$ is weakly classical. Conversely, suppose $\dscript$ is weakly classical. To show that $\qscript$ is an operator-valued measure, let $A_i$ be a sequence of mutually disjoint sets in $\ascript$. For any $\phi ,\phi '\in H$ we have that
\begin{align*}
\elbows{\qscript (\cup A_i)\phi ,\phi'}&=\elbows{\escript (\cup A_i)\phi ,\escript (\cup A_j)\phi '}\\
&=\lim _{m,n\to\infty}\elbows{\sum _{i=1}^m\escript (A_i)\phi ,\sum _{j=1}^n\escript (A_j)\phi '}\\
&=\lim _{n\to\infty}\sum _{i=1}^n\elbows{\escript (A_i)\phi ,\escript (A_i)\phi '}
  =\sum _{i=1}^\infty\elbows{\qscript (A_i)\phi ,\phi '}
\end{align*}
Hence, $\qscript$ is an operator-valued measure.
\end{proof}


\begin{thebibliography}{99}
% ref 1
\bibitem{djsu10}F.~Dowker, S.~Johnston and S.~Surya, 
On extending the quantum measure, arXiv: quant-ph (1007.2725), 2010.
% ref 2
\bibitem{dgt08}F.~Dowker and Y.~Ghazi-Tabatabai, 
Dynamical wave function collapse models in quantum measure theory, \textit{J. Phys. A} \textbf{41}, 205306  (2008).
% ref 3
\bibitem{djso10}F.~Dowker, S.~Johnston and R.~Sorkin, Hilbert spaces from path integrals, arXiv: quant-ph (1002.0589), 2010.
% ref 4
\bibitem{dut04}D.~Dutkay, Positive definite maps, representations and frames,
\textit{Rev. Math. Phys.} \textbf{16}, (2004).
% ref 5
\bibitem{gudms}S.~Gudder, Quantum measure theory, \textit{Math. Slovaca} \textbf{60} (2010), 681--700.
% ref 6
\bibitem{gud09}S.~Gudder, Quantum measure and integration theory, \textit{J. Math. Phys.} \textbf{50}
(2009), 123509.
% ref 7
\bibitem{gudamm}S.~Gudder, Examples of quantum integrals, \textit{Rep. Math. Phys.}  (to appear).
% ref 8
\bibitem{hal09}J.~J.~Halliwell, Partial decoherence of histories and the Diosi test, arXiv: quant-ph (0904.4388), 2009 and \textit{Quantum Information Processing} (to appear).
% ref 9
\bibitem{ish94}C.~Isham, Quantum logic and the histories approach to quantum theory, 
\textit{J.~Math.~Phys.} \textbf{35} (1994), 2157--2185.
% ref 10
\bibitem{joz09}R.~Jozsa, An introduction to measurement based quantum computation,
arXiv: quant-ph (0508124), 2009.
% ref 11
\bibitem{sor94}R.~Sorkin, 
Quantum mechanics as quantum measure theory, \textit{Mod. Phys. Letts.~A} \textbf{9} (1994), 3119--3127.
% ref 12
\bibitem{sor07}R.~Sorkin, 
Quantum mechanics without the wave function, \textit{J.~Phys.~A} \textbf{40} (2007), 3207--3231.

\end{thebibliography}
\end{document}